\documentclass[]{interact}

\usepackage{epstopdf}
\usepackage[caption=false]{subfig}

\usepackage[numbers,sort&compress]{natbib}
\bibpunct[, ]{[}{]}{,}{n}{,}{,}

\theoremstyle{plain}

\theoremstyle{definition}

\theoremstyle{remark}

\usepackage{float}

\usepackage{verbatim}

\begin{document}

\articletype{Application Note}

\title{A Descriptive Study of Variable Discretization and Cost-Sensitive Logistic Regression on Imbalanced Credit Data}

\author{
\name{Lili Zhang\textsuperscript{a}, Herman Ray\textsuperscript{b}\thanks{Contact Herman Ray Email: hray8@kennesaw.edu},  Jennifer Priestley\textsuperscript{b} and Soon Tan\textsuperscript{c}}
\affil{\textsuperscript{a} Analytics and Data Science Ph.D. Program, Kennesaw State University, USA; \textsuperscript{b}Analytics and Data Science Institute, Kennesaw State University, USA; \textsuperscript{c}Ermas Consulting Inc., USA}
}

\maketitle

\begin{abstract}
Training classification models on imbalanced data tends to result in bias towards the majority class. In this paper, we demonstrate how variable discretization and cost-sensitive logistic regression help mitigate this bias on an imbalanced credit scoring dataset, and further show the application of the variable discretization technique on the data from other domains, demonstrating its potential as a generic technique for classifying imbalanced data beyond credit socring. The performance measurements include ROC curves, Area under ROC Curve (AUC), Type I Error, Type II Error, accuracy, and F1 score. The results show that proper variable discretization and cost-sensitive logistic regression with the best class weights can reduce the model bias and/or variance. From the perspective of the algorithm, cost-sensitive logistic regression is beneficial for increasing the value of predictors even if they are not in their optimized forms while maintaining monotonicity. From the perspective of predictors, the variable discretization performs better than cost-sensitive logistic regression, provides more reasonable coefficient estimates for predictors which have nonlinear relationships against their empirical logit, and is robust to penalty weights on misclassifications of events and non-events determined by their apriori proportions.  
\end{abstract}


\section{Introduction} \label{introduction}
Class imbalance problems refer to a class of problems related to classifying imbalanced data where many more observations are labeled by the majority class than the minority class \cite{ali2015classification} \cite{guo2008class}. In practice, the minority class is usually the class of interest, such as fraud in the fraud detection problem \cite{phua2004minority}, malignance in the breast cancer diagnosis problem \cite{krawczyk2016evolutionary}, delinquency in the credit scoring problem \cite{brown2012experimental}, sinus bradycardia in the arrhythmia analysis \cite{guvenir1997supervised}, and poor quality in the product quality inspection \cite{cortez2009modeling}. 

However, when trained on imbalanced data, most standard statistics and machine learning models are heavily biased towards the majority class (i.e. non-events) and severely misclassify the minority class (i.e. events) \cite{visa2005issues}, caused by their assumptions of equal target class distribution \cite{japkowicz2000class} and maximizing overall accuracy \cite{provost2000machine}. Models with poor event discrimination are less useful and generate costs associated with Type II errors (money, reputation, health, etc.). 

To solve these problems more efficiently, researchers and practitioners have made efforts from various perspectives, such as data sampling \cite{krawczyk2016learning}, feature selection \cite{leevy2018survey} \cite{moayedikia2017feature}, cost-sensitive learning \cite{king2001logistic} \cite{bahnsen2014example} \cite{krawczyk2015cost}, ensemble learning \cite{collell2018simple}, and kernel-based learning \cite{ding2018kernel}, with the considerations of concrete problem characteristics. 

Previous research has not considered variable discretization as a generic technique for class imbalance problems. In this paper, we empirically explore the effects of variable discretization on classifying imbalanced data and compare it with cost-sensitive logistic regression models. Variable discretization and cost-sensitive logistic regression are studied for their high interpretability and computational efficiency. A credit scoring dataset is used in the case study. The goal is to predict the probability of a debtor's default or delinquency. The proportion of delinquency observations is only 6.68\%. We provide a detailed descriptive study on how variable discretization and cost-sensitive logistic regression help mitigate the model bias and/or variance on an imbalanced credit scoring data. The variable discretization technique is further applied on two datasets from other domains (i.e. biology, business) to demonstrate its potential for use in a wide range of fields.

The paper is structured as follows. In Section \ref{related work}, related work is reviewed. In Section \ref{data}, the data is explored and discretized. In Section \ref{modeling}, the models on the credit scoring dataset are developed, evaluated, and compared. In Section \ref{other applications}, the performance of variable discretization is examined on two datasets from other domains.  In Section \ref{discussions and conclusions}, conclusions and future work are discussed.

\section{Related Work} \label{related work}
A comprehensive review on the foundations, algorithms, and applications of imbalanced learning was conducted by He et al. in 2013 \cite{he2013imbalanced}. It summarized the previous research in five categories, including sampling methods, cost-sensitive methods, kernel-based learning methods, active learning methods, and one-class learning methods.  It also suggested to evaluate models based on both curve-based measures (e.g. ROC curve, AUC) and single-value measures (e.g. Type I Error, Type II Error, F1 score, G-mean), considering that some traditional performance measures (e.g. accuracy) did not serve as a good indicator of discrimination abilities of models \cite{rahman2013addressing}. In an imbalanced credit scoring study by Wang et al., AUC and F-measure (i.e. F1 score) were used as model performance metrics \cite{wang2015large}. 

In 2001, King proposed the weighted log-likelihood function in Eq. \ref{equation2} for the logistic regression in rare events data. Compared with the standard log-likelihood function in Eq. \ref{equation1}, Class 1 Weight ($W_1$) and Class 0 Weight ($W_0$) were added to penalize the misclassifications of events and non-events differently. $W_1$ and $W_0$ were determined by the estimated population proportion of events $\tau$ and the sample proportion of events $\bar{y}$. 
\begin{equation} \label{equation1}
\ln L(\beta | y) = \sum y_i \ln (\pi_i) + \sum (1-y_i) \ln (1-\pi_i)
\end{equation}
\begin{equation} \label{equation2}
\ln L_W(\beta | y) = W_1 \sum y_i \ln (\pi_i) + W_0 \sum (1-y_i) \ln (1-\pi_i)
\end{equation}
where
\begin{equation} \label{equation3}
\pi_i = \frac{1}{1+ \exp{- (\beta_0 + \beta_1 x_1 + ... + \beta_n x_n)}}, W_1 = \frac{\tau}{\bar{y}}, W_0 = \frac{1-\tau}{1-\bar{y}}
\end{equation}

The weighted logistic regression in Eq. \ref{equation2} is referred to as class-dependent cost-sensitive logistic regression \cite{mlr}. Bahnsen et al. proposed a different version of cost-sensitive logistic regression, called example-dependent cost-sensitive logistic regression \cite{bahnsen2014example}, where each example (i.e. observation) in the log-likelihood function was associated with a user-defined constant misclassification cost weight based on domain knowledge. Deng and Maher proposed determining each observation's cost weight by Gaussian kernel function \cite{deng1998omega} \cite{maalouf2011robust} \cite{maalouf2011kernel}, resulting in very high computational complexity $O(n^3)$ and limiting its application on big data. 

Different from cost-sensitive logistic regression which has been widely used, the variable discretization method has not been considered for addressing class imbalance problems, although it has been widely used as a domain-specific standard technique in credit scoring. This technique creates more powerful and interpretable predictors from continuous (i.e. interval) data. Dougherty et al. reviewed existing variable discretization methods, compared three of them (i.e. equal width interval, entropy-based, and purity-based) in depth on 16 datasets, and found that the global entropy-based one performed the best on average \cite{dougherty1995supervised}. For entropy-based discretization methods, the evaluation measures include: class information entropy, Gini, dissimilarity, and the Hellinger measure \cite{kotsiantis2006discretization}. For the scoring problem, one commonly used variable discretization method is called the optimal binning, which computes the cutoff points based on conditional inference trees and recursive partitioning \cite{optimal_binning_web}. 

To select powerful discretized variables, one common measurement is information value defined in Eq. \ref{equation4} \cite{hand1997statistical}, where $p_j$ is the number of non-events (i.e. non-delinquency) in the level $j$ of the variable divided by the total number of non-events, and $q_j$ is the number of events (i.e. delinquency) in the level $j$ of the variable divided by the total number of events. To interpret the information value, the following rule of thumb is proposed \cite{siddiqi2012credit} \cite{zeng2013metric}.
\begin{itemize}
\item $<0.02$: useless
\item $0.02$ to $0.1$: weak
\item $0.1$ to $0.3$: medium
\item $>0.3$: strong
\end{itemize}
\begin{equation} \label{equation4}
IV =  \sum_j (p_j - q_j)\ln (p_j/q_j)
\end{equation}
\section{Data} \label{data}
Demographic and financial information from $150,000$ borrowers is publicly available in a dataset used in a Kaggle 2011 Competition Give Me Some Credit \cite{data}. The characteristics of the individuals in the data are represented by $11$ variables, as shown in Table~\ref{table1}. The goal was to predict whether a client will experience financial distress in the next two years or not, indicated by the dependent variable $SeriousDlqin2yrs$. As shown in Table~\ref{table2}, there are $10,026$ delinquent observations and $139,937$ non-delinquent observations. The proportion of delinquencies is 6$.68\%$.  

\begin{table}[H]
\tbl{Variables for Analysis and Modeling.}
{\begin{tabular}{ l l  p{5.9cm} } \toprule
 Variable & Type & Description  \\ \midrule
 $SeriousDlqin2yrs$ & Binary & Person experienced $90$ days past due delinquency or worse \\
 $MonthlyIncome$ & Interval & Monthly income  \\
 $DebtRatio$ & Interval & Monthly debt payments, alimony, living costs divided by monthly gross income \\
 $Age$ & Interval & Age of borrower in years \\
$NumberOfDependents$ & Interval & Number of dependents in family excluding themselves (spouse, children, etc.) \\
$NumberOfOpenCreditLinesAndLoans$ & Interval & Number of open loans (installment like car loan or mortgage) and lines of credit (e.g. credit cards) \\
$NumberRealEstateLoansOrLines$ & Interval & Number of mortgage and real estate loans including home equity lines of credit \\
$RevolvingUtilizationOfUnsecuredLines$ & Interval & Total balance on credit cards and personal lines of credit except real estate and no installment debt like car loans divided by the sum of credit limits \\
$NumberOfTime30-59DaysPastDueNotWorse$ & Interval & Number of times borrower has been $30$-$59$ days past due but no worse in the last $2$ years \\
$NumberOfTime60-89DaysPastDueNotWorse$ & Interval & Number of times borrower has been $60$-$89$ days past due but no worse in the last $2$ years \\
$NumberOfTimes90DaysLate$ & Interval & Number of times borrower has been $90$ days or more past due \\ \bottomrule
\end{tabular}}
\label{table1}
\end{table}

\begin{table}[H]
\tbl{Frequency of Dependent Variable.}
{\begin{tabular}{ r r  r } \toprule
SeriousDlqin2yrs & Frequency & Percent (\%)  \\ \midrule
$1$ & $10,026$ & $6.68$ \\
$0$ & $139,974$ & $ 93.32$ \\ \bottomrule
\end{tabular}}
\label{table2}
\end{table}

There are $29,731$ observations with missing values either in the variable $MonthlyIncome$ or $NumberOfDependents$, which is $19.82\%$ of the total. These missing values are treated as follows. 

\begin{enumerate}
 \item Missing Completely at Random (MCAR) analysis is conducted, and there is no pattern existing in the missing data.  Hence, those observations are dropped to ensure the data accuracy and support the model training computation, when building the model with original variables. After dropping missing data, the proportion of delinquencies is $6.95\%$, which is very close to the original data.
 \item When building the model with discretized variables, those observations are kept by grouping the missing values separately into a level of a variable.
\end{enumerate}

\subsection{Exploratory Analysis} \label{exploratory analysis}
Because the dependent variable is binary and all independent variables are interval, the empirical logit plot is used to examine the linearity of the relationship between the dependent variable and independent variables. If the relationship is linear, it is reasonable to use the interval form of an independent variable. Otherwise, a transformation is required. Moreover, through the empirical logit plots, we can check the univariate effects, positive or negative.

The empirical logit plot is created in the following steps. 

\begin{enumerate}
  \item For each interval variable, generate percentile ranks from $1$ to $100$ \cite{rousseau2012basic}.
  \item For each rank  $i$ of each interval variable, calculate the total number of observations $N_i$, the number of delinquency observations $Y_i$, and the mean of the interval variable $\bar{x}_i$. 
  \item For each rank  $i$ of each interval variable, compute the empirical logit using the formula $elogit_i = \log (\frac{Y_i + 0.5}{N_i - Y_i + 0.5})$ \cite{donnelly2017empirical}.
  \item For each interval variable, plot the empirical logit  $elogit$ against the mean in each rank $\bar{x}$ and their linear regression line. Each point in the plot represents $N_i$ data points from the dataset by their mean. 
  \item For each interval variable, plot the empirical logit  $elogit$ against the rank $i$ and their linear regression line. Each point in the plot represents $N_i$ data points from the dataset by their rank index.
\end{enumerate}

\begin{figure}[H]
\centering
\subfloat[]{%
\resizebox*{6.5cm}{!}{\includegraphics{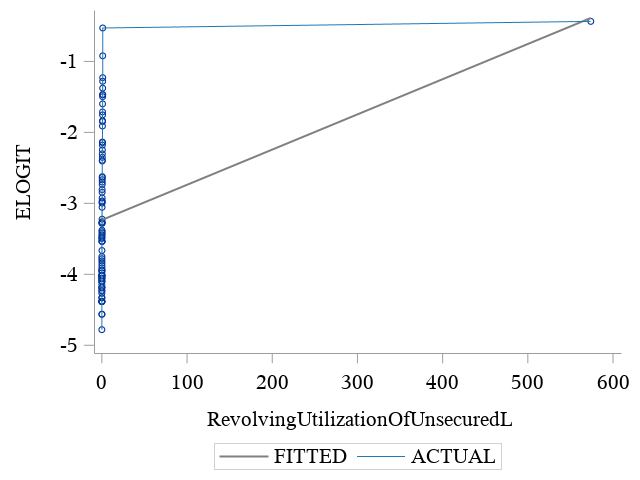} \label{figure1a}}}
\hspace{5pt}
\subfloat[]{%
\resizebox*{6.5cm}{!}{\includegraphics{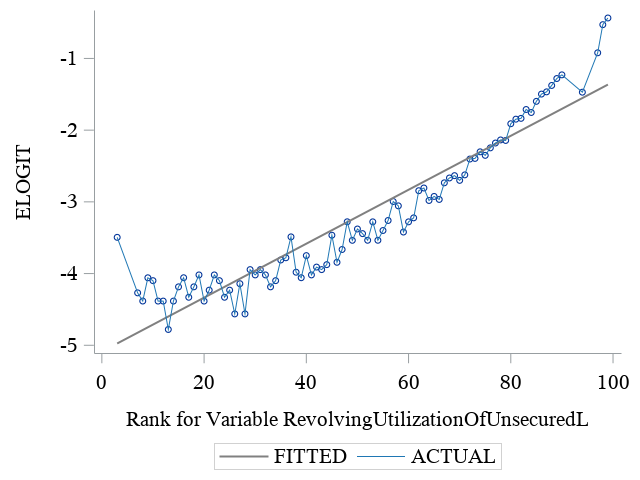} \label{figure1b}}}
\caption{Empirical Logit Plot against $RevolvingUtilizationOfUnsecuredLines$ and its Rank.} \label{figure1}
\end{figure}

For example, consider the predictor variable $RevolvingUtilizationOfUnsecuredLines$. Percentile ranks can be found in Table~\ref{elogit_data_example}. Ranks $1-8$ are merged together because their respective minimum and maximum points are the same. As shown in Figure~\ref{figure1a}, there is a nonlinear relationship between $RevolvingUtilizationOfUnsecuredLines$ and its empirical logit, mainly caused by extreme values. These extreme values in the empiricial logit plot cannot be simply removed, considering they represent several hundred data points in the dataset. However, the relationship between its rank and its empiricial logit is approximately linear as shown in Figure~\ref{figure1b}. In this case, its rank, the discretized form of its original interval values, is preferred to be used in the modeling. 

\begin{table}[H]
\tbl{Percentile Ranks of $RevolvingUtilizationOfUnsecuredLines$.}
{\begin{tabular}{ r r  r r r r r} \toprule
Rank $i$ & Min  & Max & Mean $\bar{x}_i$ & Count $N_i$ & Event  $Y_i$ & $elogit_i$ \\ \midrule
$1-8$ & $0$ & $0.000707$& $0.000034$ & $12000$ & $335$ & $-3.5488$\\
$9$ & $0.000708$ & $0.001733$ & $0.001210$ & $1501$ & $18$  &  $-4.3844$\\
$10$ & $0.001735$ & $0.002969$ & $0.002334$ & $1499$ & $25$& $-4.0574$ \\
$...$ & $...$ & $...$ & $...$ & $...$ & $...$ & $...$ \\
$99$ & $1.0062$ & $1.092954$ & $1.036357$ & $1500$ & $556$ & $-0.5290$\\
$100$ & $1.093178$ & $50708$ & $573.887190$ & $1500$ & $589$ & $-0.4358$\\
\bottomrule
\end{tabular}}
\label{elogit_data_example}
\end{table}

\subsection{Variable Discretization} \label{variable discretization}
Four variable discretization methods (i.e. distance, quantile, Gini, optimal binning) are compared. On the credit scoring dataset, the quantile discretization produces the highest AUC on the test data with the logistic regression model trained on the training data, where the ratio of training data and test data is 70\% vs. 30\%. Each variable is ranked and discretized into $20$ bins maximally based on the quantile, with the threshold value $20$ selected by the same procedure above. 

Information value is used as the measurement of the discrimination power of each individual variable after discretization, as shown in Table~\ref{information_value}. Note that for some variables, the resulting number of bins is less than 20 because the bins with non-significant differences are merged together. For the variable $MonthlyIncome$, an additional bin has been included to accomodate missing values. By following the rule suggested by Hand et. al \cite{hand1997statistical}, the variables with the information value over $0.1$ will be studied.

\begin{table}[H]
\tbl{Information Values.}
{\begin{tabular}{ l r  r } \toprule
Variables & Bins & Information Value  \\ \midrule
$RevolvingUtilizationOfUnsecuredLines$ & $19$ & $1.1635$ \\
$NumberOfTime30-59DaysPastDueNotWorse$ & $3$ & $ 0.4865$ \\ 
$NumberOfTimes90DaysLate$ & $2$ & $ 0.4842$ \\ 
$NumberOfTime60-89DaysPastDueNotWorse$ & $2$ & $ 0.2648$ \\ 
$Age$ & $20$ & $ 0.2620$ \\
$NumberOfOpenCreditLinesAndLoans$ & $15$ & $ 0.0852$ \\
$MonthlyIncome$ & $21$ & $ 0.0813$ \\
$DebtRatio$ & $20$ & $ 0.0795$ \\
$NumberOfDependents$ & $5$ & $ 0.0279$ \\
$NumberRealEstateLoansOrLines$ & $4$ & $ 0.0184$ \\
\bottomrule
\end{tabular}}
\label{information_value}
\end{table}

To prepare the discretized variables for the modeling, they are further transformed by one-hot encoding. A one-hot encoder converts a discretized variable into multiple binary dummy variables with each bin represented by one binary dummy variable \cite{onehotencoder} \cite{scikit-learn}. 

\subsection{Datasets from Other Domains} \label{datasets from other domains}

Beyond the credit scoring data, two public datasets from other domains (i.e. biology, business) are collected. They include 206 and 11 interval variables respectively, as shown in Table~\ref{datasets_info}. The goal of the arrhythmia data is to predict sinus bradycardia \cite{guvenir1997supervised}, and the goal of the wine\_quality data is to predict poor quality \cite{cortez2009modeling}. The process illustrated in Sections \ref{exploratory analysis} and \ref{variable discretization} is performed on these two datasets. Among all variable discretization methods, the optimal binning method produces the best performance. The resulting discretized variables will be modeled using logistic regression in Section \ref{other applications}.
\begin{table}[H]
\tbl{Basic Characteristics of Datasets.}
{    
\begin{tabular}{llrrrrl}
    \toprule
    Dataset & Repository & Target & Event Rate & Observations & Variables & Domain \\
    \midrule
    arrhythmia & UCI   & \multicolumn{1}{l}{06} & 5.55\% & 452   & 206C, 73N  & Biology\\
    wine\_quality & UCI   & \multicolumn{1}{l}{score$<$=4} & 3.70\% & 4898  & 11C & Business\\
    \bottomrule
    \end{tabular} }%
  \label{datasets_info}%
\end{table}%
\section{Modeling} \label{modeling}
Logistic regression and class-dependent cost-sensitive logistic regression are used as classifiers for their high interpretability. The models are evaluated by $10$-fold cross-validation. The performance measurements include ROC curve, AUC, Type I Error, Type II Error, accuracy, and F1 Score. The mean of AUCs of $10$-fold cross-validation is used to measure the model bias, while the standard deviation of AUCs of $10$-fold cross-validation is used to measure the model variance. They are reasonable measurements, considering that the model bias refers to the error introduced by approximating the true model, and the model variance refers to the amount of the change of the estimated model if using a different training dataset \cite{james2013introduction}.

To evaluate and compare the performance of variable discretization and class-dependent cost-sensitive logistic regression, the following five models are built.

\begin{figure}[h]
\centering
\subfloat[$\tau$ vs. Class Weights]{%
 \includegraphics[scale=0.2]{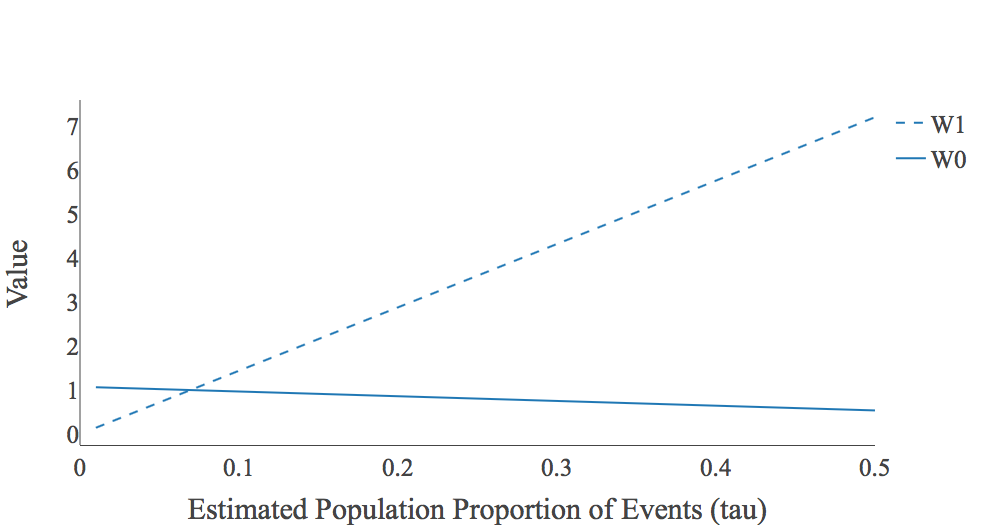}
 \label{auroc_w1_W0}}
\vspace{5pt}
\subfloat[AUROC vs. $W1$]{%
 \includegraphics[scale=0.2]{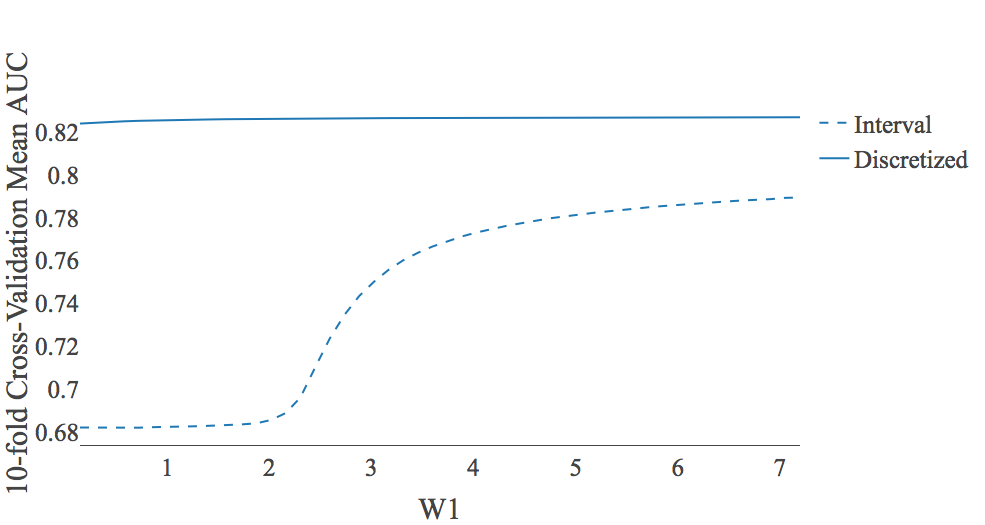}
 \label{auroc_w1}}
\caption{The Result of Tuning $\tau$.} \label{figure11}
\end{figure}

\begin{itemize}
\item Model $1$: Logistic regression model on all original interval form of independent variables in Table~\ref{table1}.
\item Model $2$: Logistic regression model on original interval form of variables with the information value over $0.1$ in Table~\ref{information_value}.
\item Model $3$: Class-dependent cost-sensitive logistic regression model on the same independent variables in Model $2$. The class weights (i.e. $W_0$, $W_1$) that produce the highest mean of AUCs of $10$-fold cross-validation are used in the modeling, indicated by the dash line in Figure~\ref{auroc_w1}. The search for the best class weights will be discussed below.
\item Model $4$: Logistic Regression model on discretized form of independent variables used in Model $2$. The discretized variables are transformed by the one-hot encoder. In total, $48$ binary dummy variables are created. 
\item Model $5$: Class-dependent cost-sensitive logistic regression model on the same discretized independent variables in Model $4$. The class weights (i.e. $W_0$, $W_1$) that produce the highest mean of AUCs of $10$-fold cross-validation are used in the modeling, indicated by the solid line in Figure~\ref{auroc_w1}.
\end{itemize}

For the class weights (i.e. $W_0$, $W_1$) in Model $3$ and  Model $5$, they are determined by the population proportion of events $\tau$ and the sample proportion of events $\bar{y}$ in Eq. \ref{equation3}. $\bar{y}$ is known from the data. $\tau$ is typically unknown and hard to obtain accurate estimation \cite{ding2016new}. Here $\tau$ is tuned as a hyperparameter from 0 to 0.5. As shown in Figure~\ref{auroc_w1_W0}, as $\tau$ increases, $W_1$ increases and $W_0$ decreases linearly. Figure~\ref{auroc_w1} shows how the mean of AUCs on the $10$-fold cross-validation changes as $W1$ increases. When modeling on interval variables in Model $3$, the best occurs at $\tau=0.5$, resulting in $W_1=7.19$ and $W_0=0.54$. The changes of the class weights have minimal influence on the modeling of discretized variables used in Model $5$, implying that good variable discretization is robust to penalty weights determined by proportions of events and non-events. Hence, for Model $5$, we take $W_1=1$ and $W_0=1$, leading Model $5$ the same as Model $4$. Because of this, we will only compare Model $4$ with other models in the following section. 

\begin{figure}[H]
\centering
\subfloat[Model 1]{%
\resizebox*{6cm}{!}{\includegraphics{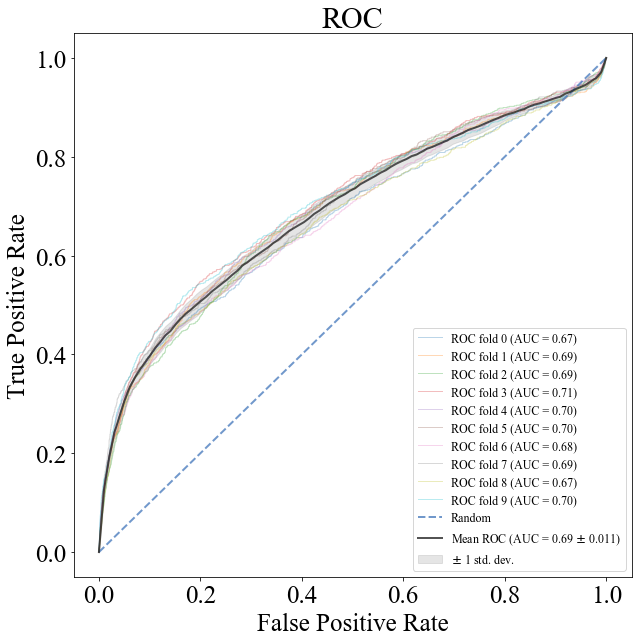} \label{figure13a}}}
\hspace{5pt}
\subfloat[Model 2]{%
\resizebox*{6cm}{!}{\includegraphics{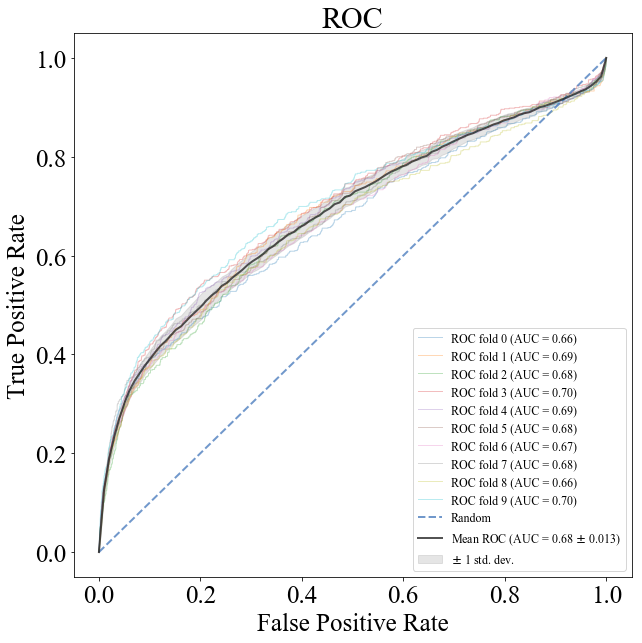} \label{figure13b}}}
\hspace{5pt}
\subfloat[Model 3]{%
\resizebox*{6cm}{!}{\includegraphics{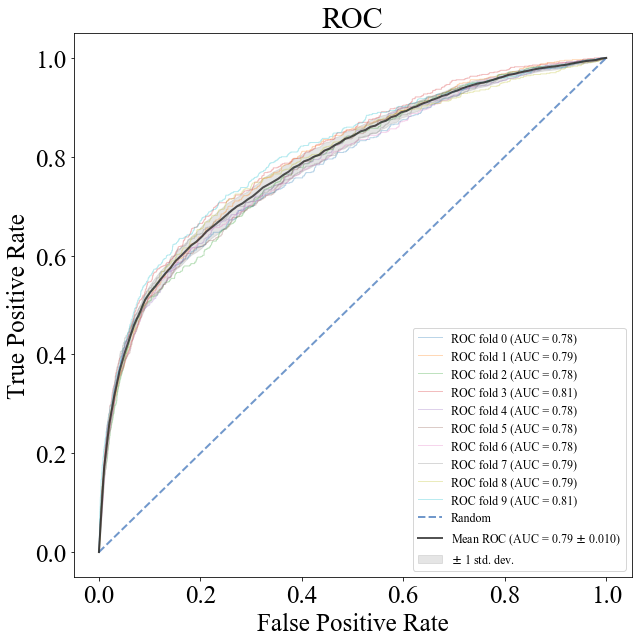} \label{figure13c}}}
\hspace{5pt}
\subfloat[Model 4]{%
\resizebox*{6cm}{!}{\includegraphics{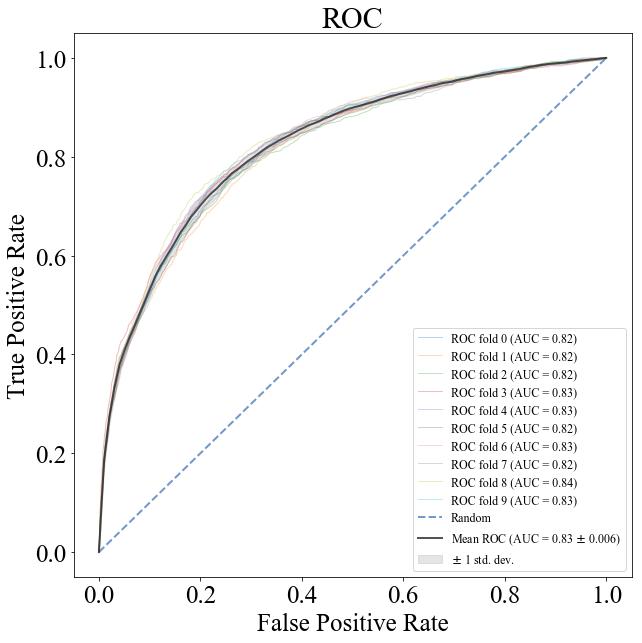} \label{figure13d}}}
\caption{10-Fold Cross-Validation ROC Curves of Credit Scoring Data.} \label{figure13}
\end{figure}

The ROC curve of each model can be found in Figure~\ref{figure13}. Model $1$ and Model $2$ have similar AUCs, indicating that the variables with the information value below $0.1$ provide minimal contribution. The ROC curves of Model $3$ and Model $4$ demonstrate stronger results than Model $2$. Moreover, for Model $4$, the ROC curves on $10$-fold cross-validation are closer to each other, indicating lower model variance. This can be further confirmed by the mean and standard deviation of AUCs on $10$-fold cross-validation in Table~\ref{table3}. Model $4$ produces the highest mean and the lowest standard deviation of AUCs, demonstrating the power of variable discretization.

\begin{table}[H]
\tbl{$10$-Fold Cross-Validation AUC of Models.}
{\begin{tabular}{ p{2cm} p{2cm} p{2cm} } \toprule
Model & Mean & Std.  \\ \midrule
Model $1$ & $0.69$ & $0.011$ \\
Model $2$ & $0.68$ & $0.013$ \\
Model $3$ & $0.79$ & $0.010$ \\
Model $4$ & $0.83$ & $0.006$ \\ \bottomrule
\end{tabular}}
\label{table3}
\end{table}

\begin{figure}[H]
\centering
\subfloat[$Age$]{%
\resizebox*{6cm}{!}{\includegraphics{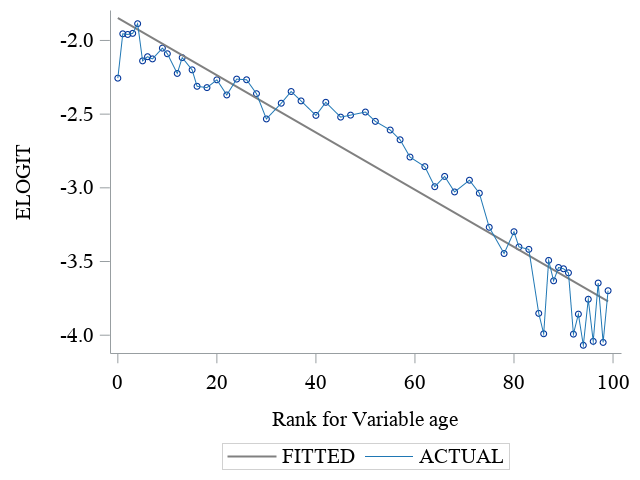}\label{age} }}
\hspace{5pt}
\subfloat[$NumberOfTime30to59DaysPastDueNotWorse$]{%
\resizebox*{6cm}{!}{\includegraphics{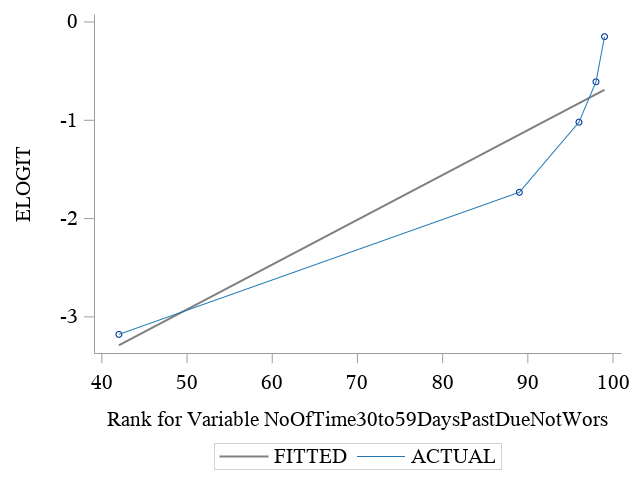}\label{30to59} }}
\hspace{5pt}
\subfloat[$NumberOfTime60to89DaysPastDueNotWorse$]{%
\resizebox*{6cm}{!}{\includegraphics{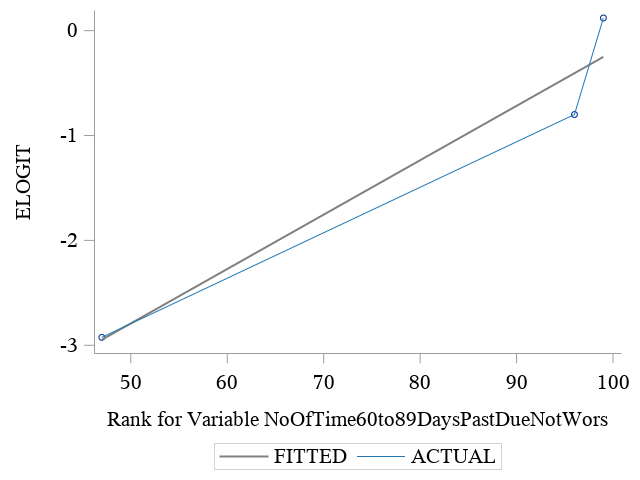} \label{60to89}}}
\hspace{5pt}
\subfloat[$NumberOfTimes90DaysLate$]{%
\resizebox*{6cm}{!}{\includegraphics{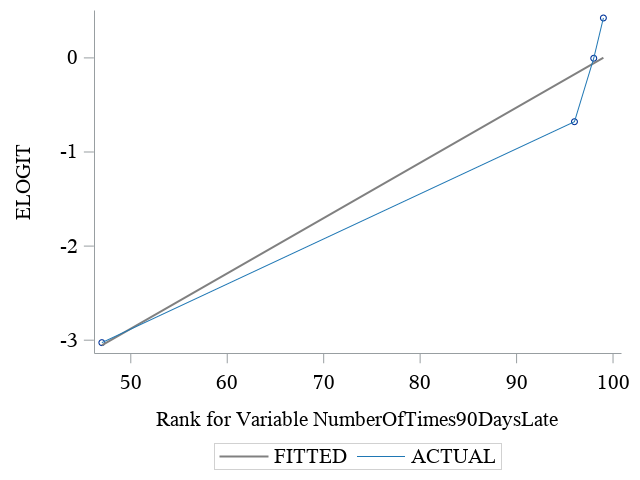}\label{90}}}
\caption{Empirical Logit Plots Against Ranks.} \label{figure_appendix}
\end{figure}

The estimated coefficients of the models are also examined. As shown in Table~\ref{estimated parameters}, Model $2$ and Model $3$ produce different estimates for every independent variable, as well as the sign of the variable $NumberOfTime60-89DaysPastDueNotWorse$. Its sign is negative in Model $2$, while its sign is positive in Model $3$. Its empirical logit plot in Figure~\ref{60to89} shows the positive relationship. Based on its variance inflation factor (VIF) in Table~\ref{VIF}, its sign change in Model $2$ is caused by its multicollinearity with the variables $NumberOfTime30-59DaysPastDueNotWorse$ and $NumberOfTimes90DaysLate$. None of them can be dropped in the modeling because of their information values presented in Table~\ref{information_value}. Model $3$ specificly guarantees a positive estimate, which is consistent with the univariate effect. For other variables, the signs of estimated parameters are consistent with their univariate effect shown in their empirical logit plots in Figures~\ref{age}, \ref{30to59}, and \ref{90}. The estimated parameters of Model $4$ are not presented here because of space limitation. Considering these dummy variables are binary indicators transformed by one-hot encoder, their estimated coefficients are more interpretable.

\begin{table}[H]
\tbl{Estimated Parameters of Model $2$ and Model $3$.}
{\begin{tabular}{l r r} \toprule
Parameter & Model $2$ Estimate & Model $3$ Estimate \\ \midrule
Intercept & $-1.45644$ & $2.69671$ \\
$RevolvingUtilizationOfUnsecuredLines$ & $-0.000048$ & $-0.000053$ \\
$NumberOfTime30-59DaysPastDueNotWorse$ & $0.50255$ & $0.67117$ \\
$NumberOfTimes90DaysLate$ & $0.45629$ & $0.79821$ \\ 
$NumberOfTime60-89DaysPastDueNotWorse$ & $-0.92206$ & $0.47276$ \\
$Age$ & $-0.02791$ & $-0.02809$ \\
\bottomrule
\end{tabular}}
\label{estimated parameters}
\end{table}

\begin{table}[H]
\tbl{VIF for $NumberOfTime60-89DaysPastDueNotWorse$ }
{\begin{tabular}{l r r} \toprule
Parameter & VIF Factor\\
$RevolvingUtilizationOfUnsecuredLines$ &  $1$\\
$NumberOfTime30-59DaysPastDueNotWorse$ & $20.5$\\
$NumberOfTimes90DaysLate$ & $20.5$\\ 
$Age$ &  $1$\\
\bottomrule
\end{tabular}}
\label{VIF}
\end{table}

Further, these models are compared based on Type I Error, Type II Error, accuracy, and F1 score on the test data after splitting the original dataset into training data (70\%) and test data (30\%), which can be found in Table~\ref{performance_measures}. The probability cutoff is chosen as the intersection point of the specificity plot and sensitivity plot, one of the most frequently used criterion \cite{habibzadeh2016determining} \cite{pramono2010prevalence}. We have the following findings.

\begin{itemize}
\item There is no improvement from Model $1$ to Model $2$, indicating that variables with information value below 0.1 provide limited contribution. 
\item Compared with Model $2$, Model $3$ decreases Type I Error by 8.23\%, decreases Type II Error by 8.3\%, increases accuracy by 8.24\%, and increases F1 score by $0.0656$, indicating the contribution of penalizing the misclassifications of events and non-events in different scales by running the class-dependent logistic regression.
\item Compared with Model $2$, Model $4$ decreases Type I Error by 11.85\%, decreases Type II Error by 11.91\%, increases accuracy by 11.85\%, and increases F1 score by $0.0946$, indicating the contribution of variable discretization. 
\item Compared with Model $3$, Model $4$ decreases Type I Error by 3.62\%, decreases Type II Error by 3.61\%, increases accuracy by 3.61\%, and increases F1 score by $0.0290$, indicating that variable discretization performs better than the inclusion of class-dependent costs in the logistic regression. 
\end{itemize}

\begin{table}[H]
\tbl{Performance Measures under the Best Probability Cutoff.}
{\begin{tabular}{ l r r r r r} \toprule
Model & Type I Error & Type II Error & Accuracy & F1 Score & Probability Cutoff  \\ \midrule
Model $1$ & $36.61\%$ & $36.54\%$ & $63.39\%$ & $0.1941$ & $0.0666$ \\
Model $2$ & $36.73\%$ & $36.74\%$ & $63.27\%$ & $0.1931$ & $0.0654$ \\
Model $3$ & $28.50\%$ & $28.44\%$ & $71.51\%$ & $0.2587$ & $0.4486$\\
Model $4$ & $24.88\%$ & $24.83\%$ & $75.12\%$ & $0.2877$ & $0.0646$ \\ \bottomrule
\end{tabular}}
\label{performance_measures}
\end{table}


\section{Application of Variable Discretization in Other Domains} \label{other applications}

To further examine the power of variable discretization, logistic regression models with original interval variables and discretized variables in the datasets arrhythmia and wine\_quality are built and compared. The original datasets are split into training data (70\%) and test data (30\%). Logistic regression models are trained on the training data and then evaluated on the test data. 

\begin{figure}[H]
\centering
\subfloat[arrhythmia]{%
 \includegraphics[scale=0.2]{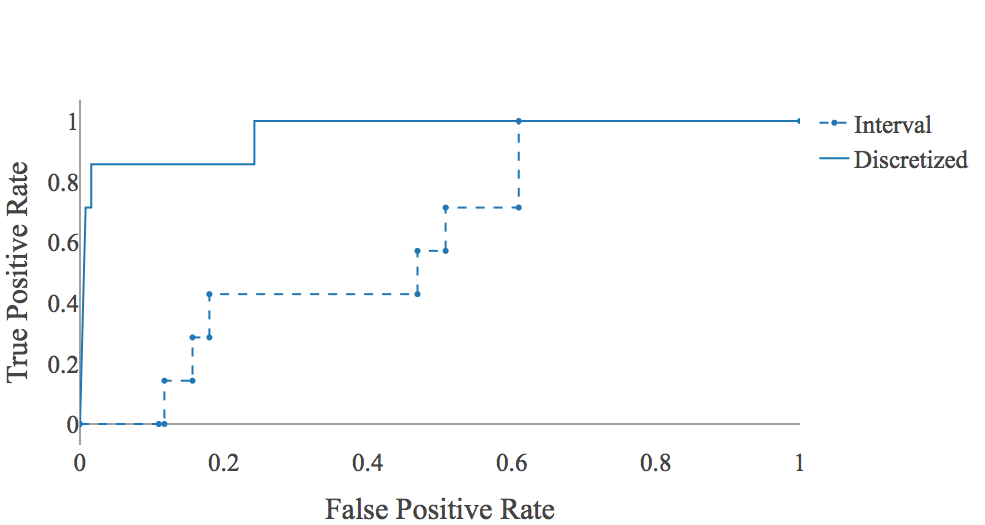}
 \label{roc_arrhythmia}}
\vspace{5pt}
\subfloat[wine\_quality]{%
 \includegraphics[scale=0.2]{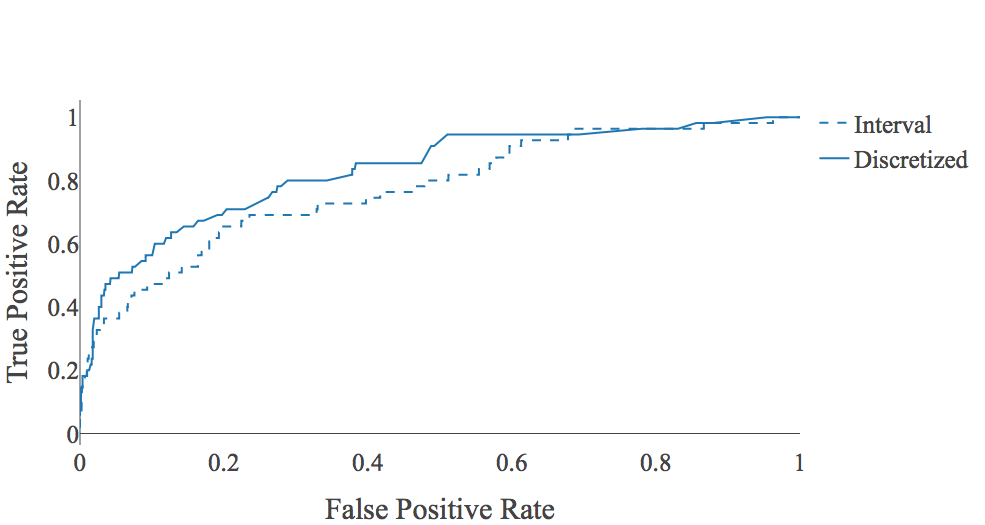}
 \label{roc_wine_quality}}
\caption{ROC Curves of Wine\_Quality and Arrhythmia Data.} \label{roc_curve}
\end{figure}

Their resulting ROC curves on the test data can be found in Figure~\ref{roc_curve}. For both datasets, the ROC curve by discretized variables moves closer to the upper-left corner than the one by interval variables. The improvement can be further checked by other performance measures (i.e. Type I Error, Type II Error, accuracy, F1 score) in Table~\ref{performance_measures_others}, where the probability cutoff is chosen as the intersection point of the sensitivity plot and specificity plot. For example, on the dataset arrhythmia, Type I Error decreases by $32.81\%$, Type II Error decreases by $27.58\%$, accuracy increases by $32.59\%$, and F1 score increases by $0.2744$. Note that the probability cutoff on this dataset is very small, but it is reasonable that some estimated probabilities are very close to 0, considering the facts that they are direct outputs of a sigmoid function ranging from 0 to 1 and target classes (i.e. non-event, event) are represented by 0 and 1 in the data.

\begin{table}[H]
\caption{The Performance of Variable Discretization on Other Datasets.}
\resizebox{\linewidth}{!} {\begin{tabular}{ l l r r r r r r} \toprule
Dataset & Model & AUC & Type I Error & Type II Error & Accuracy & F1 Score & Probability Cutoff  \\ 
\midrule
arrhythmia & Interval & $0.6216$ & $46.87\%$ & $42.86\%$ & $53.33\%$ & $0.1127$ & $3.64e-11$\\
& Discretized & $0.9603$ & $14.06\%$ & $14.28\%$ & $85.92\%$ & $0.3871$ & $2.09e-17$ \\
\midrule
wine\_quality & Interval & $0.7757$ & $30.74\%$ & $30.91\%$ & $69.25\%$ & $0.1439$ & $0.0317$ \\
& Discretized & $0.8327$ & $26.08\%$ & $25.45\%$ & $73.95\%$ & $0.1763$ & $0.0289$ \\ 
 \bottomrule
\end{tabular}}
\label{performance_measures_others}
\end{table}

\section{Discussions and Conclusions} \label{discussions and conclusions}
To improve the model performance on imbalanced data, efforts have been made from the perspective of the predictors and the modeling algorithm, respectively, in this study. Through the detailed study on the credit scoring dataset, we show that the proper variable discretization and class-dependent cost-sensitive logistic regression with the best class weights help reduce the model bias and/or variance, based on the ROC curves and AUC on $10$-fold cross-validation, Type I Error, Type II Error, accuracy, and F1 score. Moreover, class-dependent cost-sensitive logistic regression is beneficial for increasing the prediction power of predictors during the training phase even if those predictors are not transformed in their best forms and keeping the multivariate effect and univariate effect of predictors consistent. 

On the other hand, the logistic regression model with proper discretized variables performs better than class-dependent cost-sensitive logistic regression, provides more reasonable coefficient estimates, and is robust to penalty scales of misclassification costs of events and non-events determined by their proportions. This indicates that we should always discretize the variables showing nonlinear relationships against their empirical logits. 

In this study, logistic regression and its variant (i.e. class-dependent cost sensitive logistic regression) are used as classifiers. In the future, we will study the performance of variable discretization with other classifiers such as decision tree, support vector machine, and neural network. 

\bibliographystyle{tfs}
\bibliography{imbalanced_credit_data_study}

\end{document}